\documentstyle[aps,prd,preprint]{revtex}

\newcommand{\beq}{\begin{equation}}
\newcommand{\eeq}{\end{equation}}
\def\beqa{\begin{eqnarray}}
\def\eeqa{\end{eqnarray}}
\def\p{\partial}
\def\mp{m_{\rm pl}}
\def\lap{\lower.5ex\hbox{$\; \buildrel < \over \sim \;$}}
\def\gap{\lower.5ex\hbox{$\; \buildrel > \over \sim \;$}}

\def\lb{\langle}
\def\rb{\rangle}

\input psfig

\begin{document}
\tightenlines
\title{Noninteracting Dark Matter}
\author{P. J. E. Peebles\\
Institute for Advanced Study, Princeton NJ 08540\\
and\\
A. Vilenkin\\
Institute of Cosmology,\\ 
Department of Physics, Tufts University, Medford MA 02155}

\date{\today}
\maketitle
\begin{abstract}

\end{abstract}
Since an acceptable dark matter candidate may interact only weakly
with ordinary matter and radiation, it is of interest to 
consider the limiting case where the dark matter interacts only
with gravity and itself, the matter originating by the gravitational
particle production at the end 
of inflation. We use the bounds on the present 
dark mass density and the measured large-scale fluctuations 
in the thermal cosmic background radiation to constrain the 
two parameters in a self-interaction 
potential that is a sum of quadratic and quartic terms in 
a single scalar dark matter
field that is minimally coupled to gravity. In 
quintessential inflation, where the temperature 
at the end of inflation is relatively low, the field starts
acting like cold dark matter relatively late, shortly 
before the epoch of equal mass densities in 
matter and radiation. This could have observable consequences for
galaxy formation.  We respond to recent criticisms of the
quintessential inflation scenario, since these issues also 
apply to elements of the noninteracting dark matter picture.

\section{Introduction}

We have improving observational evidence\cite{observations} on 
how structure formed on the scale of galaxies and larger from  
measurements of the angular distributions of the radiation 
backgrounds, 
surveys of the spatial distributions and motions of the galaxies, 
and observations of the evolution of galaxies and the intergalactic 
medium back to redshift $z\sim 5$. It may 
prove useful to 
complement these advances with explorations of the options 
conventional physics offers for theories of structure formation. 
Here we consider the
possibility that the dark matter that is thought to 
dominate the gravitational growth of structure
interacts only with itself and
gravity, the dark matter originating by gravitational
particle production, that is, as a squeezed state, 
at the end of inflation.\footnote{Such dark matter particles might 
be called cabots, in honour of the Cabot family of Boston, who
were said to be so highly placed in society as to speak only to
God.} 

After specifying our model in Section II, we discuss the state of
the dark matter field at the end of inflation. We begin in Section~III
with the equilibrium properties of a noninteracting dark mass
scalar field
in de~Sitter spacetime. For many purposes de~Sitter spacetime is a 
useful approximation to inflation, and it has the particular advantage 
that the equilibrium statistical properties of the dark mass
field are well specified \cite{AV83,Starob,SY}.  
However, it is also of interest to analyze the effect of 
the rolling value of the Hubble parameter in a model for 
the relatively late stages of inflation. This is 
presented in Section~IV, following in part the early work by Kofman
and Linde \cite{KL}.
Section V deals with the constraints on the 
mass of the dark matter particles and on their self-coupling
from the conditions that a single scalar field has the wanted 
present mean mass density and field fluctuations that are
compatible with the large-scale anisotropy of the 
thermal background. In
quintessential inflation\cite{quint} 
this dark matter candidate
has an observationally interesting effective Jeans length.

We discuss in section VI issues of internal consistency  of the 
model. In a recent preprint 
Felder, Kofman and Linde \cite{fkl} have presented a 
stimulating list of potential problems with quintessential 
inflation and indirectly with the noninteracting dark matter model. 
We address their points in this section. 
An assessment of consistency with the full
suite of observational tests requires computations that go
well beyond our exploratory discussion; we also comment on these 
open issues in Section VI.

\section{Assumptions}

The dark matter is modeled as a single scalar field, 
$y({\bf x},t)$, that is minimally
coupled to gravity and has the self-interaction potential 
\beq
V = \lambda y^4/4 + \mu ^2y^2/2.
\label{2}
\eeq
The generalization to several field
components with $O(N)$ symmetry and the opposite sign of the
quadratic part of the potential is trivial and not interesting
for the present purpose. 

We assume that there is a time before the end of inflation 
when the Hubble parameter $H(t)$ is close enough to constant that the 
dark matter field can relax to near statistical equilibrium
between the amplitude growth driven by quantum fluctuations and
the classical slow roll due to the self-interaction potential, in an 
approximation to a de Sitter-invariant quantum state. 
For definiteness, much of our discussion of this assumption
is based on a quartic inflaton potential,
\beq
U=\lambda _\phi\phi ^4/4,
\label{Uphi}
\eeq
with a self-coupling satisfying $\lambda_\phi\lesssim 10^{-14}$, 
for consistency with the isotropy of the thermal cosmic
background radiation (the CBR). 
Kofman and Linde \cite{KL} consider the two-field model 
in Eqs~(\ref{2}) and (\ref{Uphi}) in the limit where the 
evolution of $y$ is essentially classical. Following their 
methods in part, we show in Section VI that, when 
\beq
\lambda\gg\lambda _\phi ,
\label{lambdas}
\eeq
the $y$-field energy density indeed remains subdominant 
during inflation, as required for internal consistency.
We show in Section~V that the
condition~(\ref{lambdas}) is within the ranges of values of 
$\lambda$ and $\lambda _\phi$ allowed by the CBR.

We assume the mass parameter $\mu$ is small enough that the quartic
part of the potential dominates during inflation, that $\lambda$ is
small enough that the field may be approximated as the sum of a
classical part and a free quantum field, and that 
radiative corrections to the potential in Eq.~(\ref{2}) are
negligible.  The condition for the last assumption is \cite{foot1}
\beq
{3\lambda\over{64\pi^2}}\ln{1\over{\lambda}}\ll 1.
\label{2'}
\eeq

We use a homogeneous cosmological model in which 
the field equation is 
\beq
{\p ^2y\over\p t^2} + 3{\dot a\over a}{\p y\over\p t} = 
{\nabla ^2 y\over a^2} - (\lambda y^2 + \mu ^2)y.
\label{3}
\eeq
For $\lambda y^2\ll H^2$, the oscillation frequency due to the
quartic part of the 
potential is small compared to $H$, and we can use the
slow roll approximation to the field equation,
\beq
3H\p y/\p t = -\lambda y^3.
\label{slowroll}
\eeq
This neglects the mass term, which is justified for $y^2\gg
\mu^2/\lambda$. 

We consider both the conventional model 
for reheating and the much lower temperature at the end of
inflation implied by the quintessential inflation 
model \cite{quint}. In the latter case the mass density in
interacting matter at the end of inflation is proportional to the 
fourth power of the Hubble parameter, with a coefficient that
depends on the nature of the matter and its interactions. 
We comment on the uncertainty in
this coefficient in Section~VI.

We write the Planck mass as $\mp = [3/(8\pi G)]^{1/2}$ 
in units where $\hbar = 1 = c$. 
The Hubble parameter during inflation is $H$ and the value at the
end of inflation is $H_x$.
In section V the present value of Hubble's constant
is $H_o=100h$ km~s$^{-1}$~Mpc$^{-1}$. Numerical examples assume
$h=0.7$, density parameter  
$\Omega _m=0.3$ in matter capable of clustering, 
and a cosmologically flat universe. 

\section{Statistics of the Matter Field in de Sitter Spacetime}

On scales larger than the de Sitter horizon, $H^{-1}$, the dynamics of
the field $y$ can be pictured as a ``random walk'' [superimposed on
the classical slow roll in Eq.~(\ref{slowroll})] 
in which $y$ undergoes random steps
of rms magnitude $(\delta y)_{\rm rms}=H/2\pi$ per expansion time
$\delta t=H^{-1}$, independently in each horizon-size region.  The
statistical properties of the field $y$ resulting from this random
process can be described in terms of the probability distribution
function $P(y,t)$ which satisfies the Fokker-Planck equation
\cite{AV83,Starob} 
\beq
{\partial P(y,t)\over{\partial
t}}={1\over{3H}}{\partial\over{\partial y}}[V'(y)P(y,t)] +
{H^3\over{8\pi^2}}{\partial^2 P(y,t)\over{\partial y^2}} . 
\label{FP}
\eeq
Any initial distribution, $P(y,0)$, approaches the stationary solution of
(\ref{FP}),
\beq
P_0(y)=N^{-1}\exp\left(-{8\pi^2\over{3H^4}}V(y)\right),
\label{P0}
\eeq
on the timescale \cite{SY} 
\beq
\tau _{\rm rel} \sim \lambda ^{-1/2}H^{-1}.
\label{tau}
\eeq
The coefficient $N$ in (\ref{P0}) is determined from the normalization
condition $\int_{-\infty}^{\infty}P_0(y)dy = 1$.
The stationary distribution $P_0(y)$ corresponds to the de
Sitter-invariant quantum state of the field $y$.  The statistical
properties of this state in the case of a quartic
potential $V(y)=\lambda y^4/4$ have been studied by Starobinsky and 
Yokoyama\cite{SY} (hereafter SY).

The distribution function $P_0(y)$ is not Gaussian; in particular it 
has negative excess 
kurtosis \cite{SY} $\lb\phi^4\rb/(\lb\phi^2\rb^2)-3 \approx 0.812$.
In the following 
numerical example of correlation functions of 
powers of the field, we will refer to the moments
\beqa
\lb y^2\rb &=& 0.1318H^2\lambda^{-1/2}, \quad
\lb y^4\rb - \lb y^2\rb ^2 = 0.0206H^4\lambda^{-1}, \nonumber\\
\lb y^6\rb &=& 0.01502H^6\lambda^{-3/2}, \quad
\lb y^8\rb - \lb y^4\rb ^2 = 0.00577H^8\lambda^{-2}.\label{10}\\
\lb y^{12}\rb &-& 3\lb y^4\rb\lb y^8\rb + 2\lb y^4\rb ^3
= 0.001755 H^{12}\lambda ^{-3}. \nonumber
\eeqa

SY use the Fokker-Planck formalism to compute the field
correlation function $\lb y_1y_2\rb$. Their method 
and numerical results are readily adapted to get 
the correlation function of powers of the field, 
in particular the 
correlation function of the mass density $\lambda y^4/4$.  
This is done in the Appendix; we find that for a positive integer power
$n$ of the field the equal time reduced 
correlation function
of $y^n$ at large separation $x_{12}$ is
\beq
c_n(x_{12}) = \lb y_1^ny_2^n\rb - \lb y^n\rb ^2 \propto x_{12}^{-p},
\label{11}
\eeq
where the power law index is
\beq
p = 0.178\lambda ^{1/2} \hbox{ (odd } n), \qquad 
p = 0.579\lambda ^{1/2} \hbox{ (even } n).
\label{12}
\eeq
The power law in Eq.~(\ref{11}) applies at separations large
compared to the comoving coherence length
\beq
R_c = (aH)^{-1}\exp(1/p) \sim (aH)^{-1}e^{\alpha H\tau_{rel}},
\label{rc}
\eeq
with the appropriate value of $p$ from Eq.~(\ref{12}), 
and where $\alpha$ is a number of order unity.
This expression can be interpreted to mean that in the
relaxation time  $\tau_{\rm rel}$ (Eq.~\ref{tau}) comoving
positions initially separated by the Hubble length have moved
to separation $aR_c$, with significant loss of
memory of the common initial field values at the two positions. 

As an alternative to the Fokker-Planck formalism, it may be convenient
to use a discrete version of the Langevin equation.
In an expansion time, $t=H^{-1}$, the value of the 
field $y$ 
at a fixed comoving position changes by 
the amount
\beq
y(j)\rightarrow y(j+1)= y(j) -\lambda y(j)^3/(3H^2) + 
\iota _jH/(2\pi ).
\label{6}
\eeq
The integer $j$ counts successive expansion times $H^{-1}$.
The Gaussian normal random variables $\iota _j$ have zero mean and are
statistically independent, so in particular 
$\lb\iota _j\iota _k\rb =\delta _{j,k}$.
The second term on the right-hand side of Eq.~(\ref{6}), which
represents the classical force due to the potential 
that tends to drive $y$ toward zero, 
follows from the slow roll approximation in Eq.~(\ref{slowroll}). The last
term represents the frozen quantum fluctuations added to $y$  
in the expansion time $H^{-1}$.

In the stationary de Sitter-invariant state, the mean square value of $y$ 
is
independent of time.  
This condition applied to Eq.~(\ref{6}), and keeping first powers of
the classical drift part and second powers of the fluctuating part,
yields the mean mass density in the $y$-field,  
\beq
\lb\rho _y\rb =\lambda\lb y^4\rb /4 = 3H^4/(32\pi ^2).
\label{8}
\eeq
The condition $\lb y(j)^6\rb = \lb y(j+1)^6\rb$ similarly yields
\beq
\lb y^8\rb /\lb y^4\rb ^2 = 5.
\label{variance}
\eeq
These relations agree with the Fokker-Planck
results given above, as expected.
Eq.~(\ref{variance}) says the standard deviation of the frozen field
mass fluctuations  
is $\delta\rho /\rho = 2$, with coherence length 
$R_c\sim H^{-1}\exp(1.7\lambda^{-1/2})$.

Eq.~(\ref{6}) 
may be applied as a numerical iteration prescription,
where the integers $j$ count iterations and the
numbers $\iota _j$ are drawn from a generator of
pseudo-Gaussian independent 
random numbers with zero mean and unit variance.  We use it to
calculate the two-point functions (\ref{11}) and illustrate 
the different power law indices for even and odd $n$.
The computation starts with an initial
value $y(0)$ obtained after 300 
iterations of Eq.~(\ref{6}) starting from $y=0$. Two time series,
$y_1(j)$ and $y_2(j)$, are computed from the same initial value
$y(0)$ and different sets of the $\iota _j$. This is a numerical
realization of the values of the $y$-field at successive expansion
times $H^{-1}$ at two fixed comoving positions that are
separated by the Hubble length at $j=0$. 
A set of realizations is obtained by using  
the last value of the time series $y_1$ as the initial value 
for the next realization of $y_1(j)$ and $y_2(j)$. 
The mean of $y_1^n(j)y_2(j)^n$ 
across a set of these realizations is an 
estimate of the expectation value
$\lb y_1^ny_2^n\rb$ at separation $x_{12}\propto e^j$.

Fig. 1 shows numerical results for $n=1$, 2, 3, and 4 in
Eq.~(\ref{11}). The variables are scaled to 
\beq
\hat c_n = \lambda ^{n/2}H^{-2n}c_n, \qquad
\hat x_{12} = (Hax_{12})^{\sqrt\lambda},
\label{14}
\eeq
to remove the dependence on the parameters $H$ and $\lambda$
at zero separation (Eq.~\ref{10}) and at large separation
(Eq.~\ref{12}). The heavy lines in the figure 
are the averages across realizations for two parameter choices,   
$\lambda = 0.1$ and $\lambda = 0.0001$. The latter requires a much
larger number of iterations to reach a given range of values of the
correlation functions. That reduces the number of realizations
in the computation, so the numerical noise is larger. 
The straight dotted lines are interpolations based on the SY results: the 
intercepts at zero separation are the 
one-point moments in Eq.~(\ref{10}) and the slopes are 
the power law indices 0.178 and 0.579 for the scaled 
variable $(Hax)^{\sqrt\lambda}$ (Eq.~\ref{12}).
Within the 
fluctuations from the limited number
of realizations in the averages ($1 \times 10^8$ for $\lambda =0.1$ 
and $1 \times 10^7$ for $\lambda = 1\times 10^{-4}$) the 
numerical estimates are
consistent with the expected scaling with $\lambda$.
Since the power-law asymptotics apply only for $x > R_c$, one could not
expect them to match with the one-point moments which correspond to
$ax\sim H^{-1}\ll aR_c$.  Fig.~1 indicates, however, that the
difference from a pure power law is not dramatic: the mismatch
is no more than a factor of about $2$.

The lowest curves in Fig. 1 show the mass correlation
function $\xi (x_{12}) \propto\lb y_1^4y_2^4\rb - \lb y^4\rb ^2$. At large
separation the numerical result is a factor of about 
two below the extrapolation
from zero separation. Since the reduced second moment of the mass
distribution is $\xi = 4$ at zero separation (Eq.~\ref{variance}), 
a useful approximation to the dimensionless dark mass 
correlation function is
\beq
\xi (x) = \lb\delta _1\delta _2\rb 
\sim 2(x_x/x)^\epsilon , \qquad \epsilon = 0.58\lambda ^{1/2},
\label{15}
\eeq
where $x_x$ is the comoving Hubble length at the 
end of inflation and the density contrast is
\beq
\delta =\rho /\lb\rho\rb - 1 = y^4/\lb y^4\rb - 1.
\label{16}
\eeq 

Comparison to the moments of large-scale fluctuations in
the counts of galaxies\cite{Gaz} requires higher moments of the
$y$-matter distribution; we present the example of
the three-point function. The reduced dimensionless three-point
mass correlation function is
\beqa
\xi _3 &=& \lb\delta _1\delta _2\delta _3\rb \nonumber\\
&=& \lb y_1^4y_2^4y_3^4\rb /\lb y^4\rb ^3- 
(\lb y_1^4y_2^4\rb +\lb y_2^4y_3^4\rb +\lb y_3^4y_1^4\rb )/
\lb y^4\rb ^2 + 2.
\label{17}
\eeqa
The arguments of $\xi _3$ are the lengths of the sides of the
triangle defined by the positions at which the field is evaluated. 
It is shown in the Appendix
that if the three points
define an equilateral triangle with side $x$ 
much larger than $R_c$ then the three-point function varies with $x$ as 
$\xi _3\propto \xi ^{3/2}$, where the two-point function $\xi$
(Eq.~\ref{15}) is evaluated at $x$.
The one-point moments (Eq.~\ref{10}) show that 
$\xi _3/\xi ^{3/2}=4$ at zero separation. 

Fig. 2 shows numerical
realizations of $\xi _3/\xi ^{3/2}$ for equilateral triangles.
The estimates of $\xi _3$ are based on sets of 
three independent time series $y_1(j)$, $y_2(j)$, and $y_3(j)$
with the same initial value. The values of $\lambda$, the numbers
of realizations, and the 
scaled variables are the same as for Fig. 1. The curves are 
close to the SY one-point moments at small separation, 
and, as also predicted by the Fokker-Planck method, 
at scaled separation $\hat x\gap 10$ the ratio is close to 
independent of triangle size, within the considerable noise 
from the limited number of realizations, at
\beq
\xi _3/\xi ^{3/2}\simeq 0.8
\label{18}
\eeq
This gives a useful working approximation to the normalization
of the three-point function 
on scales of interest for structure formation.\footnote{The asymptotic
value of the ratio in Eq.~(\ref{18}) can be found
by a numerical computation of the eigenfunction $\Phi_2(y)$ and
the integral $C$ in Eq.~(\ref{C}) of the Appendix.
We have not attempted this computation.}

The scaling of the $n$-point mass 
correlation functions with distance $x$ at fixed values of the 
ratios $x_{ij}/x_{kl}$ of distances among the $n$ points 
may be obtained by the following argument.
Let $\bar\delta _i=\delta\rho/\rho$ be the mass density contrast
at expansion parameter 
$a=a_i$ smoothed within a window of fixed comoving size, shape, 
and position, and let $\bar\delta _f$ be the density contrast
smoothed in the same window at a later time, 
at expansion parameter $a_f$. A realization $\delta (t)$ of
the evolution from $\delta _i$ to $\delta _f$ is a result of the 
process of freezing of quantum fluctuations and 
the classical evolution
toward $\delta = 0$. If $a_f/a_i$ is large enough that the 
epochs are separated by many relaxation times, the 
initial value $\delta _i$ is a small perturbation to this process,
and in the lowest nontrivial order in perturbation theory the 
expectation value of $\bar\delta _f$ for given initial 
value $\bar\delta _i$ is linear in $\bar\delta _i$: 
\beq
\langle\bar\delta _f\rangle _{\bar\delta _i}= T\bar\delta _i .
\label{transfer_eq}
\eeq
The transfer coefficient $T$ is a function of the window size and
shape, which we are holding fixed in comoving coordinates, 
and of the expansion factor $a_f/a_i$, which we will vary. 
Next we consider two
disjoint windows, both of fixed size, shape and position in the 
same comoving coordinate system. If the distance between the 
windows is much larger than the Hubble length at
$a_i$ the evolution of the density contrast in window~1 is
independent of the process in window~2, and it follows that 
the equal time mass two-point correlation function satisfies
\beq
\langle\bar\delta _f(1)\bar\delta _f(2)\rangle =
T^2\langle\bar\delta _i(1)\bar\delta _i(2)\rangle .
\label{AA}
\eeq
In the de Sitter equilibrium state the two-point 
mass correlation function depends only on 
the proper separation $ax$, so we can rewrite
Eq.~(\ref{AA}) as 
\beq
\xi (bx) = T(b)^2 \xi (x), \qquad b = a_f/a_i.
\eeq
This implies 
\beq
\xi\propto x^{-2\nu },\qquad T\propto b^{-\nu},
\label{AB}
\eeq
where $\nu$ is a constant that must be positive for convergence. 
That is, at large separation $x$ the correlation 
function varies as a power of $x$, 
as was shown more directly by our use of the SY analysis. 
The $n$-point function similarly satisfies
\beq
\langle\bar\delta _f(1)\ldots\bar\delta _f(n)\rangle =
T^n\langle\bar\delta _i(1)\ldots\bar\delta _i(n)\rangle ,
\eeq
from which it follows that the $n$-point function scales with
the separation $x$ at fixed ratios of separations among the $n$ points
as
\beq
\xi_n\propto x^{-n\nu },
\label{BB}
\eeq
consistent with Eq.~(\ref{18}). 

Eq.~(\ref{BB}) says the $n^{\rm th}$ central moment 
of the dark mass $M$ contained in a volume of fixed shape 
scales as $M_n=\lb (M-\lb M\rb)^n\rb\propto\sigma ^n$, where
$\sigma=(\lb (M-\lb M\rb)^2\rb)^{1/2}$ is the standard 
deviation (and assuming the integrals over the correlation 
functions converge at vanishing separation between points).
The smoothed mass fluctuations do not approach a Gaussian
as the size of the smoothing window is increased. Rather, this  
is a scale-invariant fractal in which
the mass fluctuations smoothed
and referred to the standard deviation are independent of 
the size of the smoothing window.

\section{The Dark Matter at the End of Inflation}

We have assumed so far that the evolution of the inflaton $\phi$ is
sufficiently slow that the field $y$ has enough time to settle
into its de Sitter-invariant ``equilibrium'' state.  Here we
specify the conditions of applicability of this picture under a
specific inflaton potential, and we consider the transition to
the opposite regime, which applies toward the end of inflation,
when the evolution of $y$ is almost classical.

For the inflaton potential in Eq.~(\ref{Uphi}) 
an approximate solution for the expansion history through inflation is
\cite{Lindebook}
\beq
H = H_xe^{-H_xt}, \qquad H_x={4\over 3}\lambda _\phi ^{1/2}\mp ,
\qquad \phi _x=\left( 8/3\right) ^{1/2}\mp ,
\label{Hoft}
\eeq
with expansion parameter
\beq
\log a_x/a = H/H_x - 1.
\label{axovera}
\eeq

We write the comoving length scale $x$ of field fluctuations 
frozen at time $t$, when the expansion and Hubble parameters 
are $a$ and $H$, as
\beq
Hax=1. 
\label{xoft}
\eeq
This comoving length reaches its  minimum value, $x_x$,
at the end of inflation. In Eq.~(\ref{Hoft}) inflation ends at
$t=0$, when the Hubble parameter and inflaton
field values are $H_x$ and $\phi _x$.

In the approximation of 
Eqs. (\ref{Hoft}) to (\ref{xoft}) 
the Hubble parameter $H$ when the field fluctuations on the 
comoving length scale $x$ are freezing satisfies
\beq
\log (x/x_x) \simeq \log (H_x/H) + H/H_x - 1. 
\label{simulation}
\eeq
A useful approximation at $a\ll a_x$ is
\beq
x/x_x \sim e^{H/H_x}. 
\label{xoverxx}
\eeq

The relaxation time for $y$ and the evolution time for $H$ 
are (Eqs.~\ref{tau} and~\ref{Hoft}) 
\beq
\tau _{\rm rel} \sim \lambda ^{-1/2}H^{-1},\qquad \tau _H\sim H_x^{-1}.
\label{relaxation times}
\eeq
When $\tau _{\rm rel}\ll\tau_H$ the $y$-field relaxes to
statistical equilibrium between fluctuations and classical slow roll, 
in a good approximation to the de Sitter 
equilibrium state. When $\tau _{\rm rel}\gg\tau_H$ 
the field evolution is close to classical. At the transition between 
these limiting cases, at expansion parameter $a_e$, the Hubble 
parameter is
\beq
H_e\sim \lambda ^{-1/2}H_x.
\label{xe}
\eeq

Since the $y$-field
at $a=a_e$ is close to statistical equilibrium its characteristic
value is 
$y_e\sim\lambda ^{-1/4}H_e$ (Eq.~\ref{10}). This is 
related to the value of the inflaton at $a=a_e$ by
\beq
y_e/\phi _e\sim (\lambda _\phi /\lambda )^{1/2}.
\label{eqr}
\eeq
Kofman and Linde \cite{KL} show that, if the inflaton and
the $y$-field both have quartic potentials,
the evolution in the slow roll classical approximation is 
\beq
(\lambda_\phi \phi^2)^{-1}-(\lambda y^2)^{-1} ={\rm constant}
= (\lambda_\phi \phi _x^2)^{-1}-(\lambda y_x^2)^{-1} . 
\label{KoLi}
\eeq
Eq. (\ref{eqr}) says the constant of integration is not very
important at $a=a_e$, so a good approximation to the smaller
field values at $a_e\ll a\lap a_x$ is
\beq
y(t)=(\lambda_\phi/\lambda)^{1/2}\phi(t).
\label{yphi}
\eeq

We consider first the field fluctuations frozen
at $a_e<a<a_x$, when the evolution of $y$ is close to classical. In
unit logarithmic interval of $x$ the contribution to the 
variance of the frozen dark mass
field is $\delta y_x^2=(H/2\pi )^2$. Apart from regions where $y$
happens to be close to zero, this is a small fractional perturbation
to the field value. With Eqs.~(\ref{KoLi}) 
and~(\ref{yphi}) we see that the 
field perturbation at the end of inflation is\footnote{To avoid confusion 
we remind the reader that $\delta y_x/y_x$ is the fractional 
fluctuation in the dark matter field value at the end of 
inflation. The subscript $x$ in Eq.~(\ref{dyx}) 
has nothing to do with the 
comoving length scale of the fluctuation in Eq.~(\ref{smallscalecase}).} 
\beq
{\delta y_x\over y_x} = {\delta y\over y}\left( y_x\over y\right) ^2
= {H\over 2 \pi\phi _x}\left(\phi _x\over\phi\right) ^3
\left(\lambda\over\lambda _x\right) ^{1/2}.
\label{dyx}
\eeq
Here $H$, $y$, and $\phi$ are evaluated when field fluctuations
on the scale $x$ are frozen, at $\log (x/x_x) = (\phi /\phi _x)^2$
in the approximation of Eq.~(\ref{xoverxx}). With Eq.~(\ref{Hoft}) 
we get
\beq
\delta y_x/y_x\sim 0.1\lambda ^{1/2} [\log (x/x_x)]^{-1/2}.
\label{smallscalecase}
\eeq
Since this is nearly independent of the length scale $x$
the power spectrum varies with wave number about as $k^{-3}$, 
and the mass correlation function is 
\beq
\xi\sim M\lambda\log (x_e/x), 
\label{xixp}
\eeq
where $M$ is a dimensionless constant. 

At separations larger than the field coherence
length, 
$R_c\sim (a_eH_e)^{-1}\exp(1/p)$  with $p\sim\lambda^{1/2}$ 
(Eq.~\ref{12}), Eq.~(\ref{xixp}) fails; the 
values $y_e$ may be quite different
at the two points. The classical evolution Eq.~(\ref{KoLi})
says the field values at a given comoving position
$a=a_e$ and at the end of inflation
at $a=a_x$ satisfy
\beq
{1\over y_x^2} = {1\over y_e^2} + {\lambda\over H_x^2}(1-\sqrt\lambda ).
\label{ybound}
\eeq
If $\lambda\ll 1$ this says the field value is 
$y_x\sim\pm\lambda ^{-1/2}H_x$ everywhere except near the surfaces 
where $y_e$ vanishes, and the energy density is
\beq
\rho _x\sim \lambda ^{-1}H_x^4, 
\label{rhox}
\eeq
except near the zeros of $y_e$. If the distance between zeros 
were larger than the typical motion of the dark mass after
inflation the present distribution would be close to uniform 
apart from surfaces of low density. 

We estimate the form of the mass correlation function 
at the end of inflation and for separation
$x\gg R_c$ by the argument used to obtain the 
scaling of the $n$-point mass correlation functions 
in the de~Sitter-invariant case in Eq.~(\ref{BB}). 
Consider two windows with fixed sizes and positions 
in comoving coordinates. The dark mass density contrasts
smoothed within these windows are $\bar\delta _e(1)$ and
$\bar\delta _e(2)$ at expansion parameter $a=a_e$,
when the field 
fluctuations start to depart from the de~Sitter-invariant state,
and $\bar\delta _x(1)$ and
$\bar\delta _x(2)$ at the end of inflation 
at $a=a_x$. If the comoving separation $a_ex_{12}$ of the windows
is large compared to $R_c$ at $a_e$ 
then we have, following the derivation of Eq.~(\ref{AB}),
\beq
\lb\bar\delta _x(1)\bar\delta _x2)\rb = 
T^2 \lb\bar\delta _e(1)\bar\delta _e(2)\rb 
 \sim N(x_x/x_{12})^{0.6\sqrt\lambda } \sim \xi _x(x_{12}).
\label{xix}
\eeq
The power law coefficient follows from the de~Sitter-invariant  
correlation function ~(\ref{15}). The 
transfer coefficient $T$, defined as 
in Eq.~(\ref{transfer_eq}), depends on $\lambda$. In general
it also depends on the window size and shape, but in the limit where 
the two windows are small compared to their separation Eq~(\ref{xix}) 
reduces to the two-point mass correlation function. 

Quite similarly, the three-point mass correlation function
satisfies 
\beq
\lb\bar\delta_(1)\bar\delta_x(2)\bar\delta_x(3)\rb =
T^3\lb\bar\delta_e(1)\bar\delta_e(2)\bar\delta_e(3)\rb .
\eeq  
Combined with Eq.(\ref{xix}) this implies that the
relation (\ref{18}) between the three- and two-point correlation
functions is still valid at the end of inflation.

If $\lambda$ is not much smaller than unity
the near de~Sitter evolution ends not
long before the end of inflation. In this case an estimate 
of the mass correlation function at
the end of inflation from a numerical
realization of the process is easy and useful. As in Section II,
we label successive e-foldings of the comoving Hubble length by
the integer $j$, but now $j$ decreases with time, with $j=1$ at
the end of inflation.  The values of the Hubble constant follow from
Eq~(\ref{simulation}), 
\beq
e^{j - 1} = {x_j\over x_x} = {e^{H_j/H_x - 1}\over H_j/H_x} .
\label{Hjn}
\eeq
Eq.~(\ref{KoLi}) says the 
values of the dark matter and inflation fields at successive
e-foldings are related by
\beq
{1\over\lambda }\left( {1\over y_j^2} - {1\over y_{j+1}^2}\right) =
{1\over\lambda _\phi  }\left( {1\over \phi _j^2} - {1\over \phi 
_{j+1}^2}\right).
\eeq
The result of multiplying this by $H_x^2$, rearranging, and adding
the quantum noise term $\pm H/(2\pi )$ at each 
e-folding of $x$ is 
\beq
z_j = {z _{j+1}\over [1 + G_j z_{j + 1}^2]^{1/2}} + 
{\iota _j h_j\over 2\pi },
\label{Hj}
\eeq
where the $\iota _j$ are independent Gaussian normal numbers, 
the field has been scaled to 
\beq
z =y/H_x,
\eeq
and
\beq
G_j = {2\lambda\over 3}\left( {H_x\over H_j} - {H_x\over H_{j + 1}}\right).
\eeq
The $H_j$ come from the numerical solution of
Eq.~(\ref{Hjn}). The two-point function at separation
$x/x_x=e^j$ is the mean of products of field values computed
starting from a common value $j$ time steps before 
the end of inflation. 

Figure~3 compares the mass correlation
functions at the end 
of inflation for constant Hubble parameter and for the rolling
case in Eq.~(\ref{Hoft}), for $\lambda = 0.1$.  
One sees that at relatively small separations $\xi (x)$ is
considerably flatter in the rolling $H$ case, as
expected from Eq~(\ref{xixp}). The 
values and rates of change of the correlation functions  
are roughly similar at separations $x\sim 10^{15}x_x$.
This suggests that for $\lambda =0.1$ 
the constant $N$ in Eq.~(\ref{xix}) is not greatly
different from unity.

At smaller $\lambda$ a numerical realization of the 
two-point function is difficult because
the relaxation time is long, but it is easy to get the 
one-point distribution of $y$ at the end of inflation. 
At $\lambda =0.001$ the distribution is strongly
peaked at $y\sim\pm\lambda ^{-1/2}H_x$,
as expected (Eq.~\ref{rhox}), but there is 
significant scatter from the quantum fluctuations 
appearing at $a\lap a_e$. For $\lambda =0.1$ the distribution
of field values $y$ is peaked at zero, and not greatly different
from the stationary de~Sitter case. 

In the next section we use the estimates of the 
field mass fluctuations in Eqs.~(\ref{smallscalecase})
and~(\ref{xix}) with the measured CBR anisotropy to
find bounds on the parameter $\lambda$. We will argue that the 
considerable uncertainty in $N$ translates to a relatively small 
uncertainty in the bound on $\lambda$. 

\section{The Dark Matter at the Present Epoch}

We consider three constraints: the present mean mass
density in dark matter, the small-scale cutoff in the 
gravitational growth of clustering of the $y$-matter, and the 
large-scale anisotropy of the thermal background radiation. 
Because most of our estimates are quite approximate we 
ignore most numerical factors that are of order unity. 

\subsection{The Mean Mass Density}

With the change from proper world time $t$ to  
conformal time $x^0=\int dt/a$, and the definition 
$\tilde y=ay$, the action
for the $y$-field is 
\beq
S = \int d^4x\left( {1\over 2}\tilde y_{,i}\tilde y^{,i}
-{1\over 4}\lambda\tilde y^4 - {1\over 12}a^2\tilde y^2R\right) ,
\label{19a}
\eeq
where the index is raised by the Minkowski metric tensor and we ignore 
the quadratic part of the potential. 
The Ricci tensor $R$ is on the order of the square of the Hubble
parameter $\dot a/a$ when the universe is matter-dominated, and it is
well below that when radiation-dominated. When the Ricci term
may be neglected a solution for $\tilde y$ in Minkowski
spacetime is a solution for $ay$ expressed as a function of
comoving coordinates and conformal time. The $y$-field energy
density depends on the proper time derivative
$\dot y = (\p\tilde y/\p x^0 - \tilde y\dot a)/a^2$.
If the field oscillates rapidly on the scale of the Hubble time,
either because the potential is driving rapid oscillations or the field
varies with position on scales small compared to the Hubble length, 
then $\dot y=a^{-2}\p\tilde y/\p x^0$ to good accuracy, and
the proper field energy density is well approximated
as the energy density of
$\tilde y$ in Minkowski coordinates divided by $a^4$. Since the 
former conserves energy the mean 
energy density in the $y$-field varies as
\beq
\rho _y\propto a^{-4}. 
\label{22}
\eeq
The characteristic amplitude of the field oscillations thus
scales as 
\beq
y\propto 1/a(t).
\label{23}
\eeq
Ford\cite{Ford} obtained these results
for a quartic potential when $y$ is a function only of world
time. It is an easy exercise to obtain Eqs~(\ref{22}) 
and~(\ref{23}) by extending Ford's method to
the case where the scales of length and time 
variations of $y$ both are small compared to $H^{-1}$.

The mean field value at the end of inflation is, from
Eqs.~(\ref{Hoft}) and~(\ref{yphi}), 
\beq
y_x\sim \lambda ^{-1/2}H_x,
\label{y_xvalue}
\eeq
At this field value 
the characteristic field oscillation frequency 
is $\omega\sim\lambda ^{1/2}y_x\sim H_x$.
This means the field starts oscillating at about the time 
inflation ends, so the field amplitude after inflation decays as 
$y\sim y_xa_x/a$ until the quadratic part of the potential
becomes comparable to the quartic part, at  
$y\sim y_\mu=\mu\lambda ^{-1/2}$. The expansion factor at which
this happens is 
\beq
a_\mu/a_x\sim H_x/\mu . 
\label{23a}
\eeq
The present mean mass density in the $y$-field, which 
we are assuming is comparable to the total, is
\beq
\rho _o\sim z_{\rm eq}T_o^4\sim \lambda ^{-1}
H_x^4(a_x/a_\mu )^4(a_\mu /a_o)^3.
\label{24}
\eeq
The mass density at the end of inflation is 
$\lambda y_x^4/4\sim\lambda ^{-1}H_x^4$ (Eq~\ref{y_xvalue}). 
The present CBR temperature is $T_o$, and
the redshift at equality of mass densities in
matter and radiation is
$z_{\rm eq}\simeq 3500$ (for 
$\Omega _m = 0.3$ and $h = 0.7$).

In the conventional model for reheating 
it will be sufficient to consider the case 
where the expansion is radiation-dominated at 
temperature $T_x\sim (\mp H)^{1/2}$ 
at the end of inflation. Then the expansion 
factor from the end of inflation to
the present is $a_o/a_x\sim T_x/T_o$. With Eqs.~(\ref{23a})
and~(\ref{24}) we get
\beq
\mu\sim \lambda z_{\rm eq}T_o\mp ^3/T_x^3\sim 10^5\lambda 
T_{14}^{-3}\hbox{ GeV},
\label{25}
\eeq
for temperature $T_x=10^{14}T_{14}$~GeV at the end of inflation. 
The redshift at which the dark matter starts acting as a massive
field is
\beq
a_o/a_\mu\sim\lambda z_{\rm eq}(\mp /T_x)^4\sim 10^{22}\lambda T_{14}^{-4}.
\label{26}
\eeq
If $\lambda\gg 10^{-14}$ and $T_{14}\sim 1$ 
the length scales of interest for extragalactic
astronomy appear at the Hubble
length at redshifts well below this value of  $a_o/a_\mu$, 
and the dynamical behavior of the $y$-matter is not
significantly different from the usual cold dark matter. 

In the quintessential picture the mass densities at the end of 
inflation in gravitationally produced interacting matter and
noninteracting matter are
\beq
\rho_x({\rm matter}) = CH_x^4,\qquad 
\rho _x({\rm dm})\sim\lambda ^{-1}H_x^4.
\label{mass_densities}
\eeq
The constant $C$ depends on the matter interactions, as discussed
in Section VI-A. Consistency with the standard model
for the origin of the light elements requires that the ratio 
of these mass densities, $f\sim\lambda C$,
be greater than about $f\sim 15$ at the time of light element
production. If this condition is violated the model is excluded. 
If the model is viable we can normalize the parameter constraints
to the value of $f$. The redshift at which the $y$-field starts acting
like a nonrelativistic massive field is
\beq
z_\mu = a_o/a_\mu\sim f z_{\rm eq}.
\label{27}
\eeq
The radiation temperature at the end of inflation is
$T_x\sim C^{1/4}H_x$, so the expansion factor to the present is 
$a_o/a_x\sim C^{1/4}H_x/T_o$. 
With Eq. (\ref{23a}) we have
\beq
\mu\sim \lambda C^{3/4}z_{\rm eq}T_o \sim 1\lambda C^{3/4}\hbox{ eV}.
\label{28}
\eeq

\subsection{The Effective Jeans Length}

In quintessential inflation $y$ starts acting like a massive
field at relatively low redshift (Eq.~\ref{27}), and we have to check
the effect on the gravitational growth of clustering of the
dark matter. The $y$-field fluctuations 
that appear at the Hubble length well after inflation 
have wavelengths that are much 
larger than the period of oscillation of $y$, and the 
gradient energy density $(\nabla y)^2/2$ thus is strongly 
subdominant to the kinetic and potential energy densities. We
assume dynamical evolution leaves the gradient energy subdominant.
We proceed by deriving a virial theorem valid when the field 
oscillation frequency is much larger than the Hubble parameter, 
following Ford\cite{Ford}, and use it to estimate the
effective pressure from the trace of the stress-energy tensor. 

Neglecting the cosmological expansion, the field equation is
\beq
\ddot y - \nabla ^2y + \lambda y^3 + \mu ^2y = 0.
\label{31}
\eeq
The result of multiplying this equation by $y$, averaging over space,
and averaging over a time interval that is much longer than the field
oscillation time and much shorter than the cosmological expansion time
is
\beq
\lb\dot y^2\rb = \lb (\nabla y)^2\rb + \lambda\lb y^4\rb 
+\mu ^2\lb y^2\rb .
\label{32}
\eeq
With this relation the mean energy density is
\beq
\lb\rho\rb = \lb (\nabla y)^2\rb + 3 \lambda\lb y^4\rb /4
+ \mu ^2\lb y^2\rb ,
\label{33}
\eeq
and the mean of the trace of the stress-energy tensor is
\beq
\lb T\rb = \lb\rho\rb - 3\lb p\rb = \mu ^2\lb y^2\rb ,
\label{34}
\eeq
where $p$ is the effective pressure. If the gradient energy is
subdominant we have 
\beq
{\lb p\rb\over\lb\rho\rb } = {1\over 3}{\lambda\lb y^4\rb\over
\lambda\lb y^4\rb + 4\mu ^2\lb y^2\rb /3}.
\label{35}
\eeq

Properties of $y(t)$ considered as a single oscillator with 
the equation of motion 
$\ddot y=-\lambda y^3$ are discussed by Greene, Kofman, Linde, 
and Starobinsky \cite{Greene}. The energy equation for the
oscillator is 
\beq
\dot y^2/2 + \lambda y^4/4 = \lambda y_o^4/4,
\label{36}
\eeq
where the amplitude is $y_o$, and we can use this 
to compute time averages of moments of $y$. In particular,
\beq 
\lb y^2\rb ^2/\lb y^4\rb = 48[\Gamma (3/4)/\Gamma (1/4)]^4 \simeq 0.63.
\label{37}
\eeq
When the amplitude is small enough that the 
quadratic part of the potential dominates,
\beq
\lb y^2\rb ^2/\lb y^4\rb = 2/3.
\label{38}
\eeq
In short, we can take it that $\lb y^4\rb\sim \lb y^2\rb ^2$, 
and we see from Eq.~(\ref{35}) that
$\lb p\rb /\lb\rho\rb\simeq 1/3$ until 
$\lb y^2\rb\sim\mu ^2/\lambda$, at expansion parameter $a\sim a_\mu$.
In quintessential inflation this is at redshift 
$z_\mu\sim fz_{\rm eq}$ (Eq.~\ref{27}). At $z<z_\mu$ the 
ratio varies as $\lb p\rb /\lb\rho\rb\propto a^{-3}$, because 
$\lb y^2\rb\propto a^{-3}$ for a massive nonrelativistic field.
Thus the effective velocity of sound is 
$c_s\propto a^{-3/2}\propto t^{-3/4}$ 
at $z_\mu >a>z_{\rm eq}$ and $c_s\propto t^{-1}$ 
at $z<z_{\rm eq}$. The physical Jeans length
$\sim c_st$ increases as $t^{1/4}$ at $z_\mu\gap z\gap z_{\rm eq}$
and then approaches a constant. The comoving Jeans length 
$c_st/a$ is maximum at $z_\mu$, and the maximum comoving Jeans 
length referred to the present epoch is
\beq
L _{\rm J}\sim t_\mu z_\mu /\sqrt{3}\sim 6/(f\Omega _m h^2)
\hbox{ Mpc}.
\label{39a}
\eeq
The world time at redshift $z_\mu$ is $t_\mu$, the present matter density
parameter is $\Omega _m$, and the present Hubble parameter is
$h$. On comoving scales larger than
$L_{\rm J}$ it is a good approximation to
neglect the effect of the field stress on 
the gravitational evolution of the $y$-field mass distribution.
On smaller scales 
the dynamical behavior of the field requires a more
detailed analysis than is attempted here. 

The effective Jeans length in Eq.~(\ref{39a}) is smaller than that
of a family of neutrinos with mass $\sim 30$~eV, 
and may be comparable to the mean distance between large
galaxies.

\subsection{The Large-Scale Thermal Background Anisotropy}

The parameter $\lambda$ is bounded by the effect of 
isocurvature fluctuations in the $y$-field at the end of
inflation on the angular distribution of the thermal 
background radiation (the CBR). For definiteness we present
numerical results for quintessential inflation.

This analysis of the temperature fluctuations 
on large angular scales uses the simplification 
that the radiation pressure
gradient force has little effect on the mass distribution
on the scale of the present
Hubble length, so we can imagine
that well-separated regions we see at the Hubble length evolve
as separate homogeneous cosmological 
models. In one of these homogeneous models the field value and
radiation temperature at the end of inflation are $y_x$ and
$T_x$. By repeating the computation in Section IV-A one finds
that the present dark mass density varies with these parameters and
the present temperature, $T_o$, as
\beq
\rho _o \sim\lambda ^{1/2}C^{-3/4}\mu (y_xT_o/T_x)^3. 
\label{39}
\eeq
Under the isocurvature initial conditions (and assuming the
$y$-mass density at high redshift 
is subdominant, as required for light element
production), the temperature $T_x$ is nearly 
homogeneous, the same in all 
the model universes that represent the evolution of different 
regions. Also, the present dominant mass density,
$\rho _o$, is close to homogeneous on the scale of the present Hubble
length, because pressure gradient forces cannot
generate large-scale curvature fluctuations. Thus the present
large-scale CBR anisotropy is 
\beq
{\delta T_o\over T_o}\simeq - {\delta y_x\over y_x} \simeq 
-{1\over 4}{\delta\rho _x\over\rho _x}.
\label{40}
\eeq
This relation also follows in the conventional model for reheating. 

We need expansion factors during and after inflation. 
In quintessential
inflation the ratio of the comoving Hubble length now, 
$x_o = (H_oa_o)^{-1}$, and the Hubble length $x_x$ at the end
of inflation is 
\beq
{x_o\over x_x}\sim {H_xa_x\over H_oa_o}\sim C^{-1/4}{T_o\over H_o}.
\label{46}
\eeq
The ratio of the present temperature and Hubble parameter is
\beq
T_o/H_o\sim e^{67}.
\label{TooverHo}
\eeq
When field fluctuations during inflation 
are being frozen at the comoving
scale $x_o$ of the present Hubble length
the expansion and Hubble parameters 
$a_p$ and $H_p$ satisfy 
\beq
H_pa_p\sim H_oa_o\sim C^{1/4}H_xa_xH_o/T_o.
\eeq 
These relations with Eq.~(\ref{axovera}) say
the expansion factor from $a_p$ to the end of
inflation satisfies
\beq
a_x/a_p\sim C^{-1/4}(T_o/H_o)\log a_x/a_p.
\eeq
If $C$ is on the order of unity the numerical solution is
\beq
H_p/H_x = \log a_x/a_p \simeq 72.
\label{HpoverHx}
\eeq

We have to consider two cases, where the dark matter field
departs from near statistical equilibrium
and commences classical slow roll before or after
$ax_o$ appears at the Hubble length during inflation. 
The Hubble parameter 
at the breaking of equilibrium is 
$H_e\sim\lambda ^{-1/2}H_x$ (Eq.~\ref{xe}), so a critical
parameter value is
\beq
\lambda _p =(H_x/H_p)^2\sim 1\times 10^{-4},
\label{lambdap}
\eeq
from Eq~(\ref{HpoverHx}). 
If $\lambda\ll\lambda _p$ then we observe
in the CBR anisotropy the
effect of fluctuations frozen while $y$ was behaving as a
classical field that is slightly perturbed by quantum
fluctuations. In this case 
Eqs.~(\ref{smallscalecase}), (\ref{40}) to~(\ref{TooverHo})
and~(\ref{HpoverHx}), 
with the measurement $\delta T_o/T_o\simeq 1\times 10^{-5}$,
imply 
\beq
\lambda\lap 1\times 10^{-6}.
\label{smalllambdalimit}
\eeq
This assumes $C$ is on the order of unity in Eq.~(\ref{46}),
and unless $C$ is very large its value does not significantly 
affect the bound on $\lambda$. The condition $C\sim f/\lambda$
from Eq.~(\ref{mass_densities}) with $f\sim 10$ 
requires $C\gap 10^7$. This limit is close enough 
to unity for the purpose of the numerical estimate 
in Eq.~(\ref{HpoverHx}).

The bound in Eq.~(\ref{smalllambdalimit}) differs from the
analysis of Felder, Kofman and Linde \cite{fkl} 
by the factor in Eq.~(\ref{dyx}) that takes account of the
classical decay of $y$ from $a_p$ to the end of inflation. 

In the other limiting case, 
$\lambda\gg\lambda _p$, it is convenient to use the expansion
of the observed angular distribution of the background
temperature in spherical harmonics, with coefficients
\beq
a_l^m= \int d\Omega\, Y_l^m \delta T_o/T_o.
\label{41}
\eeq
With Eq~(\ref{40}), and using the addition theorem, 
$\sum _mY_l^m(1)Y_l^{-m}(2)=(2l+1)P_l(\cos\theta _{12})/4\pi$, 
we can write the mean square value of a multipole moment as 
\beq
\left(\delta T_l\over T\right) ^2 = {l(2l+1)\over 4\pi }\lb |a_l^m|^2\rb =
{l(2l+1)\over 4\pi }\int d\Omega P_l(\cos\theta )\xi (x)/16.
\label{42}
\eeq
In this normalization, $(\delta T_l/T)^2$ is the contribution to
the variance of the sky temperature per logarithmic interval of
the spherical harmonic index $l$.  
The argument of the dark mass correlation function at the end 
of inflation is $x = 2x_o\sin\theta /2$, where the present
angular size distance $x_o$ back to high redshift is
\beq
x_o\simeq 3.2(H_oa_o)^{-1}.
\label{angularsizedistance}
\eeq
Here $\Omega _m=0.3$, we assume zero space curvature, 
and the present Hubble and expansion
parameters are $H_o$ and $a_o$.  

For the dark mass correlation
function in Eq.~(\ref{xix}) the large-scale anisotropy spectrum is 
\beq
\left(\delta T_l\over T\right) ^2 = N {l(2l+1)\over 32}
\left( x_x\over x_o\right) ^{0.6\sqrt{\lambda }}\int _{-1}^1 d\mu
P_l(\mu ) (2 - 2\mu )^{-0.3\sqrt{\lambda }}\lap 1\times 10^{-10}.
\label{spect}
\eeq
The observational bound applies at $l\lap 30$. 
We have written $x_x/x=(x_x/x_o)(x_o/x)$; the second factor
produces the last factor in the integrand. The solution to this
equation at $\lambda\ll 1$ is not relevant because the limit in
Eq.~(\ref{smalllambdalimit}) applies. In the solution at
$\lambda$ close to unity the integral is of order unity and the
value of the integral and the factor $N$ in the primeval mass
correlation function do not much matter because the sensitive
function is $(x_x/x_o)^{0.6\lambda }$. We have
\beq
\lambda \gap 0.3.
\label{largelambdalimit}
\eeq

In the conventional reheating picture $\log a_x/a_p$ is about ten 
percent smaller. To the accuracy of our estimates the constraints 
on $\lambda$ are the same. 
 
\section{DISCUSSION}

We comment first on issues of consistency and reasonableness
of the noninteracting dark matter 
picture within conventional and quintessential endings
of inflation, and then take note of some observational challenges. 

\subsection{Theoretical Issues}

For the purpose of exploring a simple example of dark matter 
that interacts only with itself and gravity
we have considered a quartic plus quadratic 
self-interaction potential.
This functional form has the arguably attractive feature that one can
choose constant coefficients that imply an interesting present mean
dark mass density and CBR anisotropy. 
We are not competent to say whether other
models for a self-interaction
potential would be more plausible from the point of view of 
fundamental physics or would produce dark matter
candidates with more interesting properties. 

In the quartic plus quadratic potential the isotropy of the CBR
requires that the dimensionless parameter $\lambda$ be in the 
narrow range
\beq
0.3\lap\lambda\lap 1,
\label{52}
\eeq
or else be quite small, 
\beq
10^{-6}\gap\lambda\gg\lambda _\phi\lap 10^{-14}.
\label{secondrange}
\eeq
The last bound is the condition on the adiabatic 
density fluctuations produced by the inflaton. 
The upper limit in Eq.~(\ref{52}) is the 
condition that the potential be safe
from significant renormalization (Eq.~\ref{2'}).
We know of no reason to 
think either range of values of $\lambda$ 
is particularly attractive within fundamental physics, though 
it is to be hoped that input from this direction eventually will
be a factor in the completion of a satisfactory theory of
structure formation.

In a recent paper Felder, Kofman and Linde\cite{fkl} 
(hereafter FKL) raise a number of issues relevant
to our analysis. They note that scalar fields 
whose particles are produced gravitationally at the end of inflation
could end up dominating the energy density, thus assuming the role of
the inflaton.  This could indeed happen if the self-couplings of
some of these fields in quartic potential models were smaller
than that of the inflaton. One sees from Eqs.~(\ref{eqr})
and~(\ref{yphi}) that the ratio of energy densities in the inflaton and
a field with self-coupling $\lambda _j$ in the quartic model is
\beq
\rho _j/\rho _\phi = \lambda _\phi/\lambda _j.
\label{condition}
\eeq
Thus consistency within this model requires
$\lambda_j \gg \lambda_\phi$.  
Since the interacting scalar
fields of the quintessential model \cite{quint} are usual spin-0
particles, like the Higgses that interact with gauge fields,
their couplings are not expected to be particularly small.
The inflaton coupling, on the other hand, is required to be small 
[Eq.(\ref{secondrange})], and there does not seem to be a problem in
meeting the conditions on $\lambda_j$.

FKL note that the $\lambda _j$ are also constrained by the 
condition that the isocurvature
fluctuations produced by the scalar fields 
not violate the CBR isotropy.
The point is valid when some of the fields represent
stable non-interacting massive particles. Indeed, we find
that the range $10^{-6}\ll\lambda _j\lap 0.3$ is
excluded for noninteracting dark matter.\footnote{Our larger 
lower bound on $\lambda _j$ seems to be mainly due to the correction
for the decay of field fluctuations in Eq~(\ref{dyx}). Our 
larger allowed range applies 
when the equilibrium between quantum excitations and classical
decay of the dark matter field persists to close to the end of
inflation, a case FKL do not consider.} 

Interacting scalar fields, such as the Higgses, also are
produced with inhomogeneous distributions at the end of
inflation. These particles interact by the usual decay,
annihilation and production processes that produce local thermal
equilibrium. In the absence of particle number conservation laws,
this local thermal equilibrium is fully 
described by a single function of position --- the  
temperature or the mass density. Apart from the adiabatic
perturbations from the inflaton 
the model predicts negligible spacetime 
curvature fluctuations at the end of inflation. This means 
the matter density fluctuations correspond to fluctuations 
in the local starting
times of cosmological expansion with universal values of the
cosmological parameters. If all matter interacts and relaxes to 
local thermal equilibrium the result at the present time is an
unacceptably homogeneous universe. If a dark matter component
does not interact with the rest of the matter, the possibility 
considered in this paper, fluctuations in composition, which 
is to say isocurvature fluctuations, remain. There may also be
adiabatic curvature fluctuations from the inflaton, of course. 
Either or both could act as seeds for structure formation.

FKL also point out that the moduli problem is more severe than in 
the conventional picture for reheating. 
Moduli are light scalar fields with only gravitational-strength
coupling to ordinary matter. They necessarily arise in 
supergravity models and superstring theories.  The 
curvature correction generally makes the effective potential of
a modulus $\chi$ during inflation different from at
late times, the minima of the two potentials tending to be 
displaced by $\Delta\chi\sim\mp$.  Thus, after
inflation one has a nearly homogeneous field $\chi$ with $\rho_\chi
\sim m_\chi^2(\Delta\chi)^2\sim m_\chi^2\mp ^2$.  The mass $m_\chi$
is determined by the supersymmetry breaking scale $\eta_{SUSY}$,
\beq
m_\chi\sim\eta_{SUSY}^2/m_{\rm pl}.
\eeq
In models with gravity-mediated supersymmetry breaking,
$\eta_{SUSY}\sim 10^{11}~$~GeV and $m_\chi\sim 10^3$~GeV.
The main problem with the moduli is 
that they may decay very late.  
With $m_\chi\sim 10^3$~GeV, the lifetime is
\beq
\tau\sim \mp^2/m_\chi^3\sim 10^5 ~s,
\label{lifetime}
\eeq
so one runs into problems with nucleosynthesis and with
photodissociation and photoproduction of light elements by the decay
products of $\chi$\cite{Ellis}.

There are two commonly discussed possible solutions 
to the moduli problem. First, 
a short period of secondary inflation could 
dilute the moduli\cite{Randall}. This naturally 
occurs in ``thermal'' inflation\cite{Lyth}, but 
seems unlikely to be effective in quintessential inflation, 
where the relative density of the moduli is much
higher than in the conventional picture, as pointed out by
FKL. Second, the minima of $V(\chi)$ during and after 
inflation may coincide due to some symmetry\cite{Dvali,Randall2,Damour}. 
In this case moduli are produced only gravitationally, like other
light scalars, and have density $\rho_\chi\sim H^4$ at the end of
inflation.  The density of matter in standard inflation is $\rho_m\sim
\mp^2 H^2$, and $\rho_\chi/\rho_m\sim H^2/\mp^2$ is small enough to solve 
the
moduli problem.  However, in quintessential inflation $\rho_m\sim H^4$
and $\rho_\chi/\rho_m\sim 1$, which is clearly too high.

There are other ways out, however.  During inflation, moduli typically
acquire masses \cite{Dvali,Dine} $m_\chi\sim \beta H$ with $\beta\gtrsim 1$,
and it is conceivable that $\beta\gg 1$ \cite{Dvali,Linmod}.  The
gravitational production of moduli would then be exponentially
suppressed.  Moduli can also develop non-perturbative potentials,
independent of supersymmetry breaking \cite{DK}.  Their mass can then
be large enough that the lifetime in 
Eq.~(\ref{lifetime}) is $\tau \ll 1 s$. Assuming coincident minima of
$V(\chi)$ during and after inflation, this would eliminate the
problem with light elements. If the moduli decay prior to
baryogenesis, then the relaxation to thermal 
equilibrium eliminates the isocurvature perturbation
associated with the inhomogeneous primeval local ratio 
of entropy density to the number density of modulus
quanta. Otherwise, the isocurvature perturbation survives in the
form of an inhomogeneous ratio of the baryon and photon number
densities and can later play a role in structure formation.

Another possibility is that the supersymmetry breaking scale could be
much smaller than $10^{11}$~GeV; in some models it can be as low as
the electroweak scale \cite{ADD}.  
The moduli would then be very light, their lifetime
would be much greater than the present age of the
universe, and problems with nucleosynthesis would not arise.
(Here we still assume coincident minima and also that the moduli
develop masses $\sim H$ during inflation).

There is a related issue for gravitinos. The lower energy
density at the end of inflation in the quintessential 
picture reduces thermal gravitino production
but increases the problem of gravitational production, as 
FKL note. Apart from potential energy terms the gravitino obeys 
a conformally invariant field equation \cite{Grishchuk}. If during 
inflation
the gravitino had a substantial effective mass, 
the gravitino energy density from gravitational particle 
production could be
unacceptable in quintessential inflation. We are not aware that 
the effective mass has to be this large, however. 

The moduli and gravitino quanta are close approximations
to the proposed noninteracting dark matter, but with potentially 
unacceptable production of mass density. This is a challenge, 
but within the not inconsiderable uncertainties of a 
supersymmetric theory that has not yet been fully
specified the challenge does not seem serious enough to 
discourage further exploration of the picture. 

We note finally that the density of matter produced in
quintessential inflation can be substantially higher than our original
estimate in Ref.\cite{quint}.  The quanta of
spin-0 fields $\chi_j$ produced due to the change in the expansion law
at the end of inflation have wavelength $\sim H_x^{-1}$ and energy
density $\rho_j^{(s)}\sim H_x^4$.  In addition to these
short-wavelength quanta, there is also a nearly homogeneous component
of $\chi_j$.  The classical solution (\ref{yphi}) always applies at
the end of inflation, and the energy of the homogeneous component 
follows as in Eq.~(\ref{mass_densities}),
\beq
\rho_j^{(h)}\sim(\lambda_\phi^2/\lambda_j)\mp^4\sim\lambda_j^{-1} H_x^4,
\eeq
where $H_x\sim \lambda_\phi^{1/2}\mp$ is the expansion rate at the end
of inflation.  We see that $\rho_j^{(h)}\gg\rho_j^{(s)}$ for small
$\lambda_j$.  For ordinary (interacting) matter fields, the
homogeneous component decays and thermalizes, just like the
short-wavelength quanta. We note also that this
enhancement of the energy density in the homogeneous component
disappears if the fields $\chi_j$ develop masses $m_j\sim H$ during
inflation.  In this case the original estimate of \cite{quint},
$\rho_j\sim H_x^4$, is still valid, for $\lambda_j < 1$,
because the quartic term is small compared to the quadratic part 
of the potential.

\subsection{Observational Issues}

Previous examples of noninteracting dark 
matter \cite{LindeMuk}, \cite{Peeb} postulate a quadratic
self-interaction potential with a time-variable mass. 
These are CDM models, where the CDM is defined as matter 
that has had negligible pressure since appearance at the 
Hubble length of all length scales of astrophysical interest. 
The two-field CDM model in \cite{Peeb} could be adjusted to fit the
spectrum  $\delta T_l/T$ of angular fluctuations of the thermal 
background, to the accuracy of the measurements then available, 
but at $l\sim 50$ the model power is a factor of two above the
new Python\cite{Python} result. Bridging this 
gap would require even greater contrivance. This 
isocurvature CDM model thus does not seem 
to be viable, but it does offer a useful reference for 
a first assessment of observational tests of the quartic plus
quadratic self-interaction model. 

In the isocurvature CDM  model the primeval distribution of
the radiation is much smoother than that of the CDM because the
mean mass density in the CDM is much smaller. With 
the quartic potential considered here the dark matter
approximates a fluid with the equation of state of the 
radiation until the redshift approaches $z_{\rm eq}$
(Eq.~\ref{27}). This means
a given primeval isocurvature fluctuation 
$\delta\rho _y/\rho _y$ in the dark matter
distribution produces a much larger primeval perturbation to the
distribution of the radiation than for CDM. 
The effect on the fluctuation
spectrum of the CBR on angular
scales below the limit of application of Eq.~(\ref{spect}) 
remains to be computed.

In the isocurvature CDM model the primeval mass distribution is 
$\rho\propto y ({\bf x})^2$, with $y$ a random Gaussian
process, and the non-Gaussian mass density
fluctuations in this model may violate the measured 
skewness and kurtosis of the large-scale fluctuations of 
galaxy counts\cite{Gaz}. The numerical realizations in
Fig.~2 indicate that in the quartic potential
model the skewness of the primeval mass distribution 
is a factor of about 3 smaller than in the case
$\rho\propto y ^2$. 
Again, the effect on the observational
constraint may be worth investigating. 

In quintessential inflation the
dynamical behavior of the $y$-field would be an important factor
in the formation of the first generations of structure 
(Eq.~\ref{39a}). Khlebinkov and Tkachev\cite{Tkachev} present 
numerical simulations of the three-dimensional behavior of a
scalar field with quartic self-interaction potential, in
connection with the dynamics of reheating in conventional
inflation. They find a marked tendency for the growth of
field fluctuations on ever smaller scales. The qualitative effect
is easily understood: where the field value $|y|$ is larger than
average the field oscillation frequency is larger, and the
frequency differences at different positions produce growing field
gradients. Our preliminary numerical experiments in one space
dimension plus time suggest the growing field gradients produce
energy fluxes that redistribute energy and the space distribution
of field oscillation frequencies. This tends to unwind the field
gradients through substantial parts of space, while leaving
localized regions with relatively large field gradient energy 
density.  
The effect is curious enough that further study would be at
least intellectually interesting, and it is 
conceivable that it adds an interesting 
complication to the observed growth of structure on the scale 
of galaxies. 

Finally, we return to our opening remark, that it may  be
useful to complement the growing observational basis for a theory
of structure formation with surveys of the 
options available within conventional physics. 
It is natural to consider simplest possibilities first, 
but prudent to bear in mind the possibility of complications. An
example is the evidence that the density parameter in
matter capable of clustering on scales much smaller
than the Hubble length is significantly below unity\cite{Omega}.
If this were valid it would mean that the large-scale nature of
the universe is not as simple as we could imagine within conventional 
physical ideas, a result that could hardly be described as 
surprising in light of the ample opportunities
for complexity in the physical universe. We arrived at a simple
model for structure formation, adiabatic CDM, fifteen years
ago, after abandoning a few even simpler 
ideas. In the present paper we have considered a class of dark matter
candidates whose properties could be considerably more
complicated than cold dark matter, and the observational 
predictions much harder to work out. On the other hand,
the picture does not seem unnatural: possibly some of the fields
gravitationally produced by 
inflation are unable to thermalize with ordinary matter; 
perhaps such fields end up with an observationally 
interesting dark mass density; perhaps their irregular
primeval distributions are a significant factor in seeding the
gravitational growth of structure. 

\section{Acknowledgments}

We thank Lev Kofman for an advance copy of his recent paper with
Felder and Linde \cite{fkl}; it helped improve our paper.
We have also benefitted from 
discussions with Gia Dvali, Massimo Giovannini, Andrei
Gruzinov, Lev Kofman, Andrei Linde, Alison Peebles, David Spergel, 
and Paul Steinhardt.  This research was 
supported in part at the Institute for Advanced Study 
by the Alfred P. Sloan Foundation and at Tufts University by the
National Science Foundation.

\section{Appendix A. Calculation of two- and three-point functions}

To calculate the two- and three-point correlation functions in de
Sitter space, it will be convenient to introduce the conditional
probability $\Pi(y,t|y_0,t_0)$ for the field to have value $y$ at time
$t$, given that it had value $y_0$ at time $t_0$ at the same co-moving
position.  This probability satisfies the Fokker-Planck equation
(\ref{FP}) with the initial condition
\beq
\Pi(y,t_0|y_0,t_0)=\delta(y-y_0),
\eeq
and can be expressed as \cite{Risken}
\beq
\Pi(y,t|y_0,t_0)=e^{-v(y)+v(y_0)}\sum_{k=0}^\infty
\Phi_k(y)\Phi_k(y_0) e^{-\Lambda_k(t-t_0)}.
\label{Pi}
\eeq
Here,
\beq
v(y)\equiv {4\pi^2\over{3H^4}}V(y),
\eeq
and $\Phi_k(y)$ form a complete orthonormal set of eigenfunctions of
the equation
\beq
\left[-{\partial^2\over{\partial y^2}}+v'(y)^2-v''(y)\right]\Phi_k(y) =
{8\pi^2\Lambda_k \over{H^3}}\Phi_k(y).
\eeq
The eigenvalues $\Lambda_k$ are non-negative, with the smallest
eigenvalue $\Lambda_0=0$ corresponding to
\beq
\Phi_0(y)=N^{-1/2}e^{-v(y)} , 
\label{Phi0}
\eeq
with
\beq
N=\int_{-\infty}^\infty e^{-2v(y)}dy.
\eeq
Note that $\Phi_0(y)$ is related to the ``equilibrium'' distribution
function (\ref{P0}),
\beq
P_0(y)=\Phi_0^2(y).
\label{equil}
\eeq
we shall assume that the eigenfunctions are ordered so that
$0<\Lambda_1 < \Lambda_2 < ...~$.  For a purely quartic potential
\cite{SY},
\beq
\Lambda_1 = 0.0889\lambda^{1/2}H, ~~~~~~ \Lambda_2 = 0.289\lambda^{1/2}H.
\eeq

Let us now consider two co-moving positions ${\bf x_1}$ and ${\bf
x_2}$ separated by a distance $ax=a|{\bf x_1-x_2}|\gg H^{-1}$ at some
moment $t=0$.  These positions were within each other's horizons prior
to the time
\beq
t_a=-H^{-1}\ln (Hax).
\label{ta}
\eeq
For $t<t_a$, the values of $y$ at ${\bf x_1}$ and ${\bf x_2}$ are
essentially the same.  Hence, the probability for $y$ to take a value
$y_1$ at ${\bf x_1}$ and a value $y_2$ at ${\bf x_2}$ at time $t=0$
can be expressed as \cite{SY}
\beq
P_2[y_1({\bf x_1}),y_2({\bf x_2})]=\int\Pi(y_1,0|y_a,t_a)
\Pi(y_2,0|y_a,t_a) P_0(y_a)dy_a,
\label{p2}
\eeq
where $P_0(y)$ is the equilibrium distribution (\ref{equil}).  The
equal-time two-point correlation function for $y^n$ is given by
\beqa
\lb y^n({\bf x_1},0)y^n({\bf x_2},0)\rb &\equiv & \lb y_1^n y_2^n\rb
=\int_{-\infty}^\infty dy_1 dy_2 y_1^ny_2^n P_2[y_1({\bf
x_1}),y_2({\bf x_2})]\nonumber\\
 &=& N^{-1}\sum_{k=0}^\infty A_{(n)k}^2
(Hax)^{-2\Lambda_k/H},
\label{series}
\eeqa
where
\beq
A_{(n)k}=\int_{-\infty}^\infty y^ne^{-v(y)}\Phi_k(y)dy
\label{ak}
\eeq
and we have used Eqs.~(\ref{Pi}), (\ref{ta}), (\ref{p2}) and the
orthonormality of the functions $\Phi_k(y)$.  The mode functions
$\Phi_k(y)$ are even functions of $y$ for $k$ even and odd functions
of $y$ for $k$ odd ($\Phi_k(y)$ has $k$ nodes), and it is clear from
Eq.~(\ref{ak}) that
$A_{(n)k}$ are non-zero only when $n$ and $k$ are both even or both odd.

The first term in the series (\ref{series}) is 
\beq
N^{-1}A_{(n)0}^2=\lb y^n\rb^2,
\eeq
and thus the reduced two-point function,
\beq
c_n(x)=\lb y_1^ny_2^n\rb-\lb y^n\rb^2,
\eeq
is given by the same series starting with $k=1$.  The asymptotic
behavior of $c_n(x)$ at large $x$ is determined by the first term in
that series.  For odd values of $n$,
\beq
c_n(x)\approx N^{-1}A_{(n)1}^2(Hax)^{-2\Lambda_1/H}.
\eeq
For even values of $n$, $A_{(n)1}=0$ and
\beq
c_n(x)\approx N^{-1}A_{(n)2}^2(Hax)^{-2\Lambda_2/H}.
\label{xi2}
\eeq
Eqs.~(\ref{11}), (\ref{12}) of Section III follow immediately from these
relations.

Quite similarly, the three-point function can be expressed as
\beq
\lb y_1^4y_2^4y_3^4\rb =\int dy_1dy_2dy_3 y_1^4y_2^4y_3^4 
P_3[y_1({\bf x_1}),y_2({\bf x_2}),y_3({\bf x_3})].
\label{tpf}
\eeq
Assuming that ${\bf x_1},{\bf x_2},{\bf x_3}$ form an equilateral
triangle of side $ax$, the probability distribution $P_3$ can be
written as
\beq
P_3[y_1,y_2,y_3]=\int\Pi(y_1,0|y_a,t_a)
\Pi(y_2,0|y_a,t_a)\Pi(y_3,0|y_a,t_a) P_0(y_a)dy_a,
\label{p3}
\eeq
Combined with equations (\ref{tpf}),(\ref{Pi}),(\ref{ta}), this gives
\beq
\lb y_1^4y_2^4y_3^4\rb=N^{-1}\sum_{k,l,m=0}^\infty A_{(4)k}A_{(4)l}
A_{(4)m} (Hax)^{-(\Lambda_1 +\Lambda_2 +\Lambda_3)/H}
\int\Phi_k(y)\Phi_l(y)\Phi_m(y) e^{v(y)}dy.
\eeq

Once again, it is easily understood that the reduced three-point
function (\ref{17}) is given by the same series with the summation
starting at $k,l,m=2$.  The leading term at large $x$ is
\beq
\xi_3(x)\approx N^{-1}A_{(4)2}^3 C(Hax)^{-3\Lambda_2/H},
\label{xi3}
\eeq
where
\beq
C=\int_{-\infty}^\infty \Phi_2^3(y)e^{v(y)}dy.
\label{C}
\eeq
>From (\ref{xi2}) and (\ref{xi3}) we find that at large $x$
\beq
\xi_3(x)/\xi_2^{3/2}(x)\approx N^{1/2}C,
\eeq
independent of $x$.

\begin{figure}
\centerline{\psfig{file=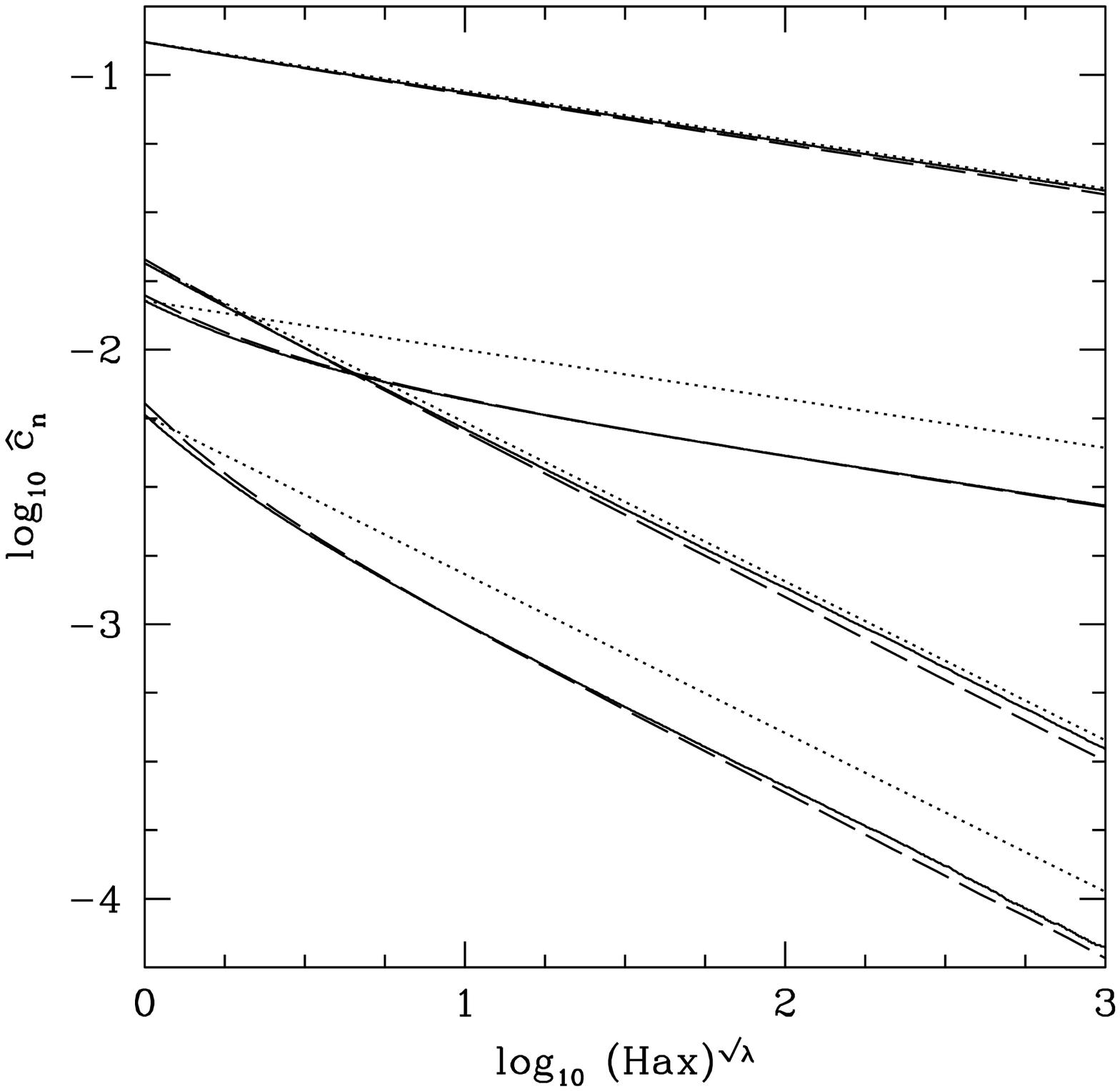,width=3truein,clip=}}
\caption{Numerical realizations of the reduced two-point correlation
functions of powers $n$ of the field $y$ (Eq.~\ref{11}), for
$n=1$, 2, 3, and 4 from top to 
bottom at the left-hand side of the figure. The dotted lines have the
intercept at zero separation and the slope at large separation of the
Fokker-Planck approximation (Eqs.~\ref{10} to~\ref{12}). The
curves are averages across 
numerical realizations of the random process for 
$\lambda = 0.1$ (dashed lines) and
$\lambda = 1\times 10^{-4}$ (solid lines). The correlation
functions and the separation are scaled to remove the dependence
on $\lambda$ and $H$ (Eq.~\ref{14}).}
\end{figure}

\begin{figure}
\centerline{\psfig{file=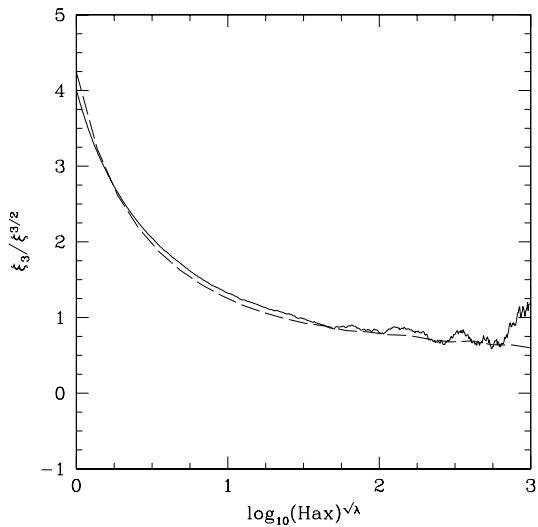,width=3truein,clip=}}
\caption{The reduced three-point correlation function $\xi _3$ for
equilateral triangles scaled by the two-point function $\xi$
evaluated at the length of the side of the triangle for 
$\lambda = 1\times 10^{-4}$ (solid line) and
$\lambda = 0.1$ (dashed line). In the Fokker-Planck 
approximation the ratio
plotted on the ordinate is constant on scales much larger than
the coherence length $aR_c$.} 
\end{figure}

\begin{figure}
\centerline{\psfig{file=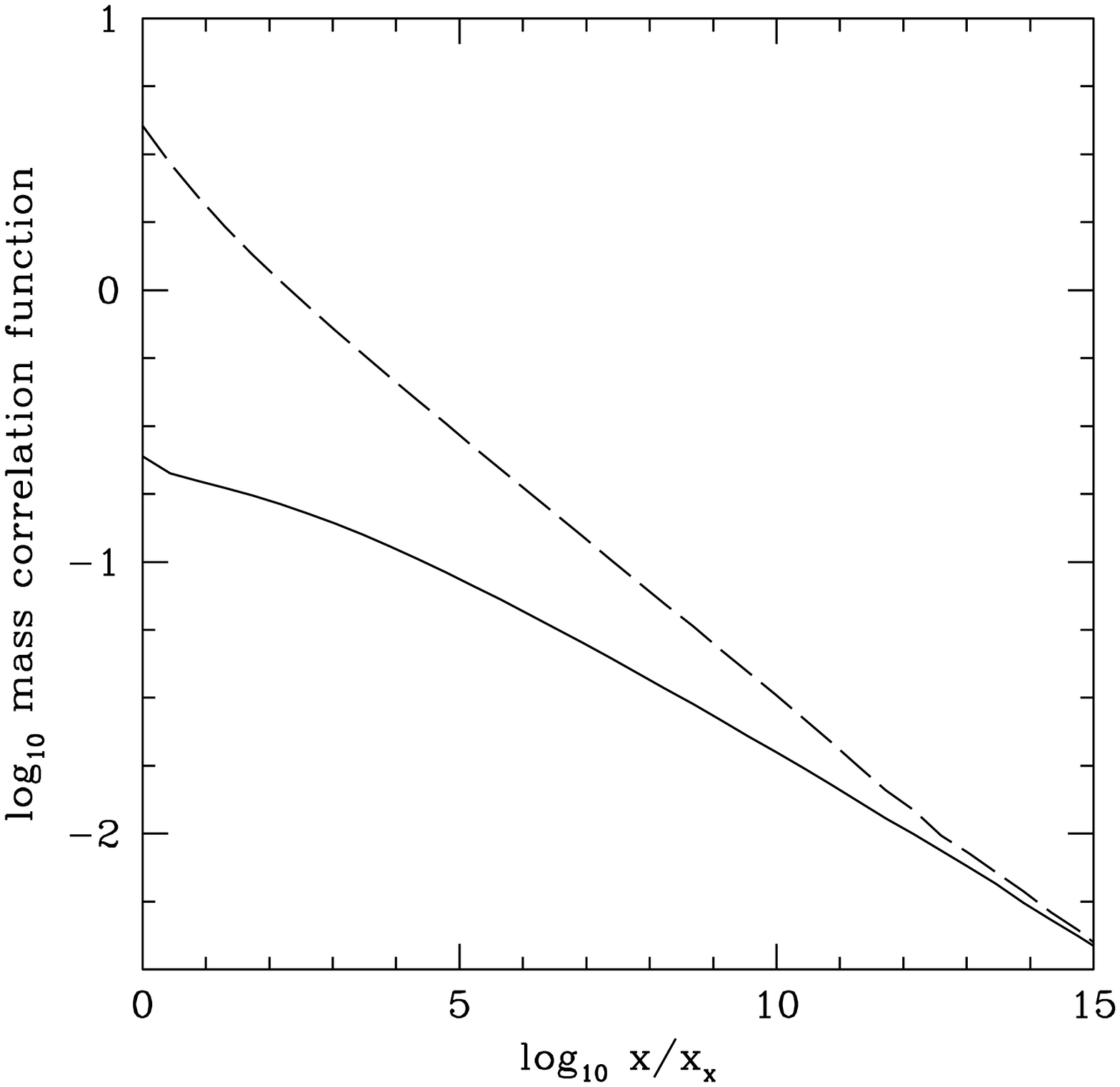,width=3truein,clip=}}
\caption{Numerical realizations of dark mass
two-point correlation functions at the 
end of inflation for the de~Sitter equilibrium state (dashed
line) and the rolling Hubble parameter in Eqs~(\ref{Hjn}) 
and~(\ref{Hj}). The
self-coupling parameter in the dark mass potential 
is $\lambda =0.1$}
\end{figure}


\begin{thebibliography}{}
\bibitem{foot1} The one-loop effective potential for the model
(\ref{2}) in de Sitter space was calculated in Ref.\cite{EP}.  The
condition (\ref{2'}) is obtained by requiring that the one-loop
correction is negligible for $y^2\lesssim \lb y^2\rb\sim
\lambda^{-1/2}H^2$, where the expectation value $\lb y^2\rb$ is taken in
the de Sitter-invariant state (see section III).
\end{thebibliography}
\end{document}